# ESTIMATION OF EFFORT IN SOFTWARE COST ANALYSIS FOR HETEROGENOUS DATASET USING FUZZY ANALOGY

S.Malathi

Research Scholar, Dept of CSE,
Sathyabama University,
Chennai, Tamilnadu, India
malathi_raghu@hotmail.com

Dr.S.Sridhar

Research Supervisor,Dept of CSE & IT,
Sathyabama University,
Chennai, Tamilnadu, India
drssridhar@yahoo.com

*Abstract*— **One of the significant objectives of software engineering community is to use effective and useful models for precise calculation of effort in software cost estimation. The existing techniques cannot handle the dataset having categorical variables efficiently including the commonly used analogy method. Also, the project attributes of cost estimation are measured in terms of linguistic values whose imprecision leads to confusion and ambiguity while explaining the process. There are no definite set of models which can efficiently handle the dataset having categorical variables and endure the major hindrances such as imprecision and uncertainty without taking the classical intervals and numeric value approaches. In this paper, a new approach based on fuzzy logic, linguistic quantifiers and analogy based reasoning is proposed to enhance the performance of the effort estimation in software projects dealing with numerical and categorical data. The performance of this proposed method illustrates that there is a realistic validation of the results while using historical heterogeneous dataset. The results were analyzed using the Mean Magnitude Relative Error (MMRE) and indicates that the proposed method can produce more explicable results than the methods which are in vogue**

## I. INTRODUCTION

Software cost estimation has gained tremendous importance in the last two decades due to its imperative necessity for efficient effort estimation in software analysis. In general, effort estimation for software projects is categorized as algorithmic and non algorithmic models [1]. Algorithmic estimation deals with the application of mathematical computation method while Non algorithmic estimation is essentially based on machine learning techniques. Software cost estimation by analogy is one of the most conspicuous machine learning techniques and is basically a form of Case-Based Reasoning [2]. Estimation by analogy is based on the assumption that similar software projects have similar costs. However, the technique needs improvement especially while handling the categorical variables.

Fuzzy logic cost estimation models [3] are more suitable for projects with indistinct and imprecise information. The advantage of this method is that it interprets the linguistic values very much similar to the human way of interpretation. However, this method is not able to overcome the imprecision and uncertainty problem in an efficient manner. The proposed method resourcefully estimates the software effort using Fuzzy analogy technique based on reasoning by analogy and fuzzy logic.

The paper is divided into 5 sections as follows. Section 2 discusses the related work. The key features of the Fuzzy Analogy approach are presented in section 3. In section 4, an explorative analysis is conducted for validating the proposed method and based on the results; a refined Fuzzy Analogy approach with the performance outcome is illustrated in section 5. The conclusion of the findings is dealt in Section 6.

## II. RELATED WORK

Effort estimation during the initial stages of project development is invariably essential for the software industry to cope with the unrelenting and competitive demands of today's world. The estimation should also be accurate, reliable and precise to meet the growing demands of the industry. Keung [4] demonstrates that the estimation by analogy is a viable alternative to predict accuracy and flexibility where the prediction of effort is done by selecting a set of completed projects which are akin to the new projects. Hasan Al-Sakran [5] has highlighted that retrieval of similar projects from the dataset can be effectively done by an improved CBR integrated with different methods. Recently, a new method has been proposed [6] to improve Analogy based software estimation by conducting empirical experiments with tools such as ESTOR and ANGEL.

A new framework has been elucidated [7], based on fuzzy logic, for estimation of effort during the initial stage itself, especially for projects representing linguistic variables. A transparent and improved Fuzzy logic based framework [8] is proposed for effectively dealing with the imprecision and uncertainty problem. The Gaussian MFs [9] have been used in the fuzzy framework, which show good results while handling the imprecision in inputs. The ability of this method to adapt itself to the varying environment in as much as its efficient handling of the inherent imprecision and uncertainty problem makes it a valid choice for representing fuzzy sets.

A multi agent system has been employed [10] to deal with the characteristics of the team members in a fuzzy system. Many studies have been carried out [11] which utilize the fuzzy systems to deal with the ambiguous and linguistic inputs of software cost estimation. In [12], it is noted that homogeneous dataset results in better and more accurate effort estimates while the irrelevant and disordered dataset results in lesser accuracy in effort estimation. Wei Lin Du et al. [13] proposed a methodology combining the neuro-fuzzy technique and SEER-SEM that can function with various algorithmic models.

## III.    PROPOSED WORK

### A.    Analogy

The basic idea of prediction of effort in cost estimation by analogy [14] is that projects having similar features such as size and complexity will be similar with respect to project effort. The method gains its importance since the estimate is based on actual project experience.

### B.    Fuzzy Logic

Fuzzy logic is based on the human behaviour and reasoning. It is similar to fuzzy set theory and used in cases where decision making is difficult. A Fuzzy set can be defined by assigning a value for an individual in the universe of discourse between the two boundaries that is represented by a membership function.

$$A = \int_x \mu_A(x)/x \qquad (1)$$

Where x is an element in X and $\mu_A(x)$ is a membership function. A Fuzzy set is represented by a membership function that has grades between the interval [0, 1] called grade membership function.

### C.    Fuzzy Analogy

Fuzzy analogy is the fuzzification of classical analogy procedure. It comprises of three steps. 1) Identification of cases 2) Retrieval of similar cases and 3) Case adaptation. Each step is the fuzzification of its equivalent classical analogy procedure.

### 1) Identification of cases:    The main objective of Fuzzy analogy method is to effectively deal with the categorical data. In identification of cases, each project is indicated by a set of selected attributes which can be measured by numerical or

categorical values. These values will be represented by fuzzy sets. In the case of numerical value $x_0$, its fuzzification will be done by the membership function which takes the value of $1$ when $x$ is equal to $x_0$ and $0$ otherwise

For categorical values, it is supposed to have $M$ attributes and for each attribute $M_j$, a measure with linguistic value is defined ($A_k^j$). Each linguistic value, $A_k^j$, is represented by a fuzzy set with a membership function ($\mu_{A_k^j}$). It is preferable that these fuzzy sets satisfy the normal condition. The use of fuzzy sets to represent categorical data, such as 'very low' and 'low', is similar to the way in which humans interpret these values and consequently it allows to deal with the vagueness, imprecision and uncertainty in the case identification step.

### 2) Retrieval of Cases:    This step is based on the selection of software project similarity measure. In retrieval of cases, a set of candidate measures is proposed for selecting software project similarity. These measures assess the overall similarity of two projects $P_1$ and $P_2$, $d(P_1,P_2)$ by combining all the individual similarities of $P_1$ and $P_2$ associated with the various linguistic variables $V_j$ describing the project $P_1$ and $P_2$, $d_{V_j}(P_1,P_2)$. After an axiomatic validation of some proposed candidate measures for the individual distances $d_{V_j}(P_1,P_2)$, two measures have been retained [15].

$$d_{V_j}(P_1,P_2)=\begin{cases} \max_k \min(\mu_{A_k^j}(P_1),\mu_{A_k^j}(P_2)) \\ \text{max–min aggregation} \\ \sum_k \mu_{A_k^j}(P_1)\times\mu_{A_k^j}(P_2) \\ \text{sum–product aggregation} \end{cases} \qquad (2)$$

Where $A_k^j$ is the fuzzy set associated with $V_j$ and $\mu_{A_k^j}$ which are the membership functions representing fuzzy sets $A_k^j$.

$$\text{Effort} = A * (\text{SIZE})^{B+0.01*\sum_{i=1}^{N} d_i} * \prod \text{EM}_i \qquad (3)$$

Where $A$ and $B$ are constants, d is the distance and EM effort multipliers. By using the above formula the effort is estimated.

*3) Case Adaptation:* The objective of this step is to derive an estimate for the new project by using the known effort values of similar projects. In the proposed method, all the projects in the data set are used to develop an estimate of the new project. Each historical project will contribute to the calculation of the effort of the new project according to its degree of similarity with this project.

## IV. RESULTS AND DISCUSSION

The historical heterogeneous dataset used in this study is the Desharnais, Nasa 60 and Nasa 93 dataset published in PRedictOR Models in Software Engineering (PROMISE) [18]. The proposed work is implemented by using the default packages of JAVA Net beans. Table 1, summarizes the number of projects collected under each dataset with the actual average effort compared with the estimated average effort using fuzzy analogy method.

**TABLE 1.** COMPARISON OF ACTUAL AVG. EFFORT WITH ESTIMATED AVG.EFFORT.

| Dataset | No. of Projects | No. of features | Actual Avg.Effort | Estimated Avg.Effort |
|---------|------|------|------|------|
| Nasa60 | 60 | 2 | 406.413 | 359.324 |
| Nasa93 | 93 | 2 | 734.031 | 530.148 |
| Desharnais | 77 | 2 | 5046.308 | 4786.311 |

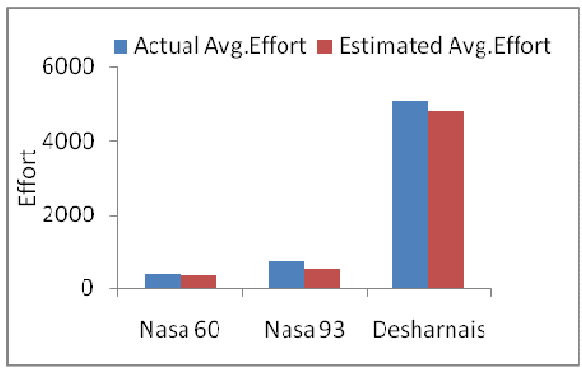

Figure.1. Comparison of Actual Avg Effort and Estimated Avg.Effort

Fig.1 indicates the comparative performance of actual average and estimated average effort for the 3 dataset.

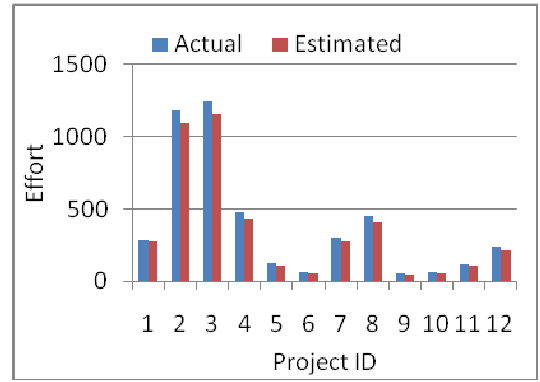

Figure.2. Comparative Results of actual and estimated effort with the Nasa 60 dataset

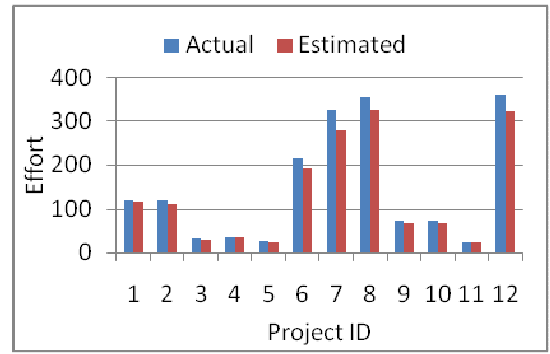

Figure.3. Comparative Results of actual and estimated effort with the Nasa 93 dataset

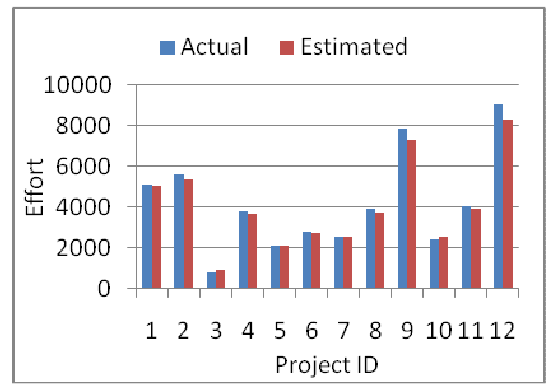

Figure.4. Comparative Results of actual and estimated effort with the Desharnais dataset

From Fig.2-4, it is inferred that the proposed method is very efficient with less effort value compared to the actual effort present in the existing three dataset.

## V. PERFORMANCE ANALYSIS

A common criterion for the evaluation of effort estimation models is the Mean Magnitude of relative Error (MMRE). The MRE and MMRE can be measured by employing the following formulae

$$MRE = |act_i - est_i| / act_i \qquad (4)$$

$$MMRE(\%) = \frac{1}{n}\sum_{i=1}^{n} MRE * 100 \qquad (5)$$

Where the $act_i$ is the actual effort, $est_i$ is the estimated effort and N is the no of cases. The comparison of proposed method with the existing method based on MMRE measure is tabulated in Table 2.

**TABLE 2.** COMPARISON OF MMRE IN PERCENTAGE WITH THE EXISTING METHODS.

| Dataset | Nasa60 | Nasa93 | Desharnais |
|---|---|---|---|
| Proposed Method | 5.15 | 6.95 | 4.98 |
| Analogy with Fuzzy Number | 33.37 | 28.55 | 26.89 |
| Fuzzy method | 32.651 | 54.81 | 30.6 |

The MMRE measure for Nasa 60,Nasa 93 and Desharnais dataset of the proposed method is compared to the existing method [3][16] [17].

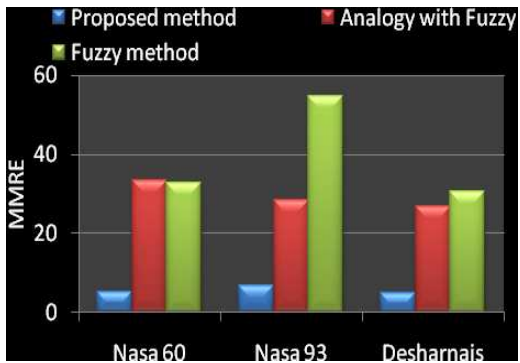

Figure.5. Comparison Performance of MMRE (%)

Fig.5 clearly depicts that the MMRE value for the heterogeneous dataset is very low compared to the different existing methods, thereby proving that the proposed method is very efficient.

## VI. CONCLUSIONS

The existing techniques for estimation of effort in software cost analysis are not able to handle the categorical variables in as much as they could not overcome the imprecision and uncertainty problem in an efficient manner. Fuzzy analogy has been developed subsequently to address these issues. However, the results are not very effective while handling the categorical data and necessitate improvement. Fuzzy analogy based on reasoning by analogy, fuzzy logic and linguistic quantifiers has been utilized for enhancing the performance as well as to overcome the imprecision and uncertainty. In the proposed method, both categorical and numerical data are represented by fuzzy sets. The salient benefit of this method is that it can overcome the imprecision and uncertainty problem to a considerable extent while describing the software project. The results also clearly indicate that proposed method effectively estimates the effort for the historical heterogeneous project datasets.

However, the existing methods and present research work deals only with the project characteristics for effort estimation but the important attributes such as team characteristics have been neglected. Therefore, the future research warrants a pragmatic approach to include the team member characteristics to evaluate the project effort in a resourceful manner.